\documentclass[12pt,preprint,showpacs,preprintnumbers,nobibnotes]{revtex4}
\usepackage{amssymb}
\usepackage{amsmath}
\usepackage{graphicx}
\usepackage{dcolumn}
\usepackage{bm}
\usepackage{epsfig}
\usepackage{array}
\usepackage{cases}
\newcommand{\PreserveBackslash}[1]{\let\temp=\\#1\let\\=\temp}
\newcolumntype{C}[1]{>{\PreserveBackslash\centering}p{#1}}
\newcolumntype{R}[1]{>{\PreserveBackslash\raggedleft}p{#1}}
\newcolumntype{L}[1]{>{\PreserveBackslash\raggedright}p{#1}}

\begin{document}

\preprint{ }

\title{Effects of interfaces on dynamics in micro-fluidic devices: slip-boundaries' impact on rotation characteristics of polar liquid film motors}
\author{Zhong-Qiang Liu$^{1}$}
\altaffiliation{Corresponding author. E-mail address:
phyzhqliu@163.com}
\author{Su-Rong Jiang$^{1}$}
\author{Tamar A. Yinnon$^{2}$}
\author{Xiang-Mu Kong$^{1}$}
\author{Ying-Jun Li$^{3}$}
\affiliation{$^{1}$Department of Physics, Qufu Normal University,
Qufu 273165, China\\
$^{2}$K. Kalia, D.N. Kikar, Jordan 90666,
Israel\\
$^{3}$SMCE, China University of Mining and Technology, Beijing
100083, China}
\date{\today }

\begin{abstract}

Slip-boundary effects on the polar liquid film motor (PLFM) -- a
novel micro-fluidic device with important implications for advancing
knowledge on liquid micro-film's structure, dynamics, modeling and
technology -- are studied. We develop a mathematical model, under
slip boundary conditions, describing electro-hydro-dynamical
rotations in the PLFMs induced either by direct current (DC) or
alternating current (AC) fields. Our main results are: (i) rotation
characteristics depend on the ratio $k=l_{s}/D$ ($l_{s}$ denotes the
slip length, resulting from the interface's impact on the structure
of the liquid and $D$ denotes the film's diameter). (ii) As $k$
($k>-1/2$) increases: (a) PLFMs subsequently exhibit rotation
characteristics under ``negative-", ``no-", ``partial-" and
``perfect-" slip boundary conditions; (b) the maximum value of the
linear velocity of the steady rotating liquid film increases and its
location approaches the film's border; (c) the decay of the angular
velocities' dependency on the distance from the center of the film
slows down, resulting in a macroscopic flow near the boundary. (iii)
In addition to $k$, the rotation characteristics of the AC PLFM
depend on the magnitudes, the frequencies, and the phase difference
of the AC fields. (iv) Our analytical derived rotation speed
distributions are consistent with the existing experimental ones.\\


\end{abstract}

\maketitle

\section{INTRODUCTION}

Exploring the physical properties, in particular hydrodynamical
characteristics, of liquid films under electric, magnetic or
electro-magnetic fields, is an endeavor of major theoretical and
technological importance for advancements in physics, biophysics and
engineering\cite{Jones2002,Nelson2003,SquiresandQuake2005,ChangandYeo2010}.
Currently, electro-hydro-dynamical (EHD) motions of liquid crystal
films are studied intensively and their unique properties are
applied widely in
industry\cite{Sonin1998,Faetti1983,Morrisetal1990,Dayaetal1997}.

The recently invented suspended polar liquid film motor (PLFM)
provides a good platform for studying hydrodynamics of
two-dimensional fluids, as well as micro-structures of different
polar liquid films, including liquid crystal
films\cite{Amjadietal2009,Shirsavaretal2011,Shirsavaretal2012,Amjadietal2013}.
This motor consists of a quasi-two-dimensional electrolysis cell in
an external in-plane electric field -- see Fig. 1. PLFMs operating
with different electric fields have important implications for
micro- motors, mixers, or
washers\cite{Amjadietal2009,Shirsavaretal2011,Liuetal2011,Liuetal2012,Liuetal2013,Liuetal2012-2};
moreover, recent experiments show that a rotating suspended liquid
film can be used as an electric generator\cite{Amjadietal2013-2}.

In previous studies (Liu et al. Phys. Rev. E 2011, 2012), we
developed models for the PLFMs which enabled quantitative and
qualitative explanations for numerous experimental results, e.g.,
its rotation direction, threshold fields for onset of EHD and the
distribution of its angular velocity\cite{Liuetal2011,Liuetal2012}.

The impact of the PLFM's film boundaries on its EHD motions hitherto
has not been addressed, i.e., all previous models assumed no-slip
hydrodynamic boundary conditions at film
borders\cite{Liuetal2011,Liuetal2012,Liuetal2013,Liuetal2012-2,Amjadietal2013,Shiryaevaetal2009,GrosuandBologa2010}.
However, experiments show EHD motions in PLFMs, near their films'
borders, depend on polar liquid type. For example: a macroscopic
observable almost static region exists near the boundary of the
rotating N-(4-methoxybenzylidene)-4-butylaniline liquid crystal
film\cite{Shirsavaretal2012}; for the rotating 2,5-Hexadione film,
the rotation's linear velocity dependence on the distance from the
center of the film decreases slowly to zero on approaching the
border\cite{Shirsavaretal2011}, a large linear velocity appears at
the border of the rotating Benzonitrile
film\cite{Shirsavaretal2011}. Moreover, recent experiments and
simulations show: negative slippage exists in hydrophilic
micro-channels\cite{Song2012} and on interfaces with a strong
solid-fluid attraction\cite{Zhang2004}; numerous no-slip and
partial-slip phenomena of polar liquids (e.g., water) on various
solid interfaces were reported\cite{Lauga2007}; large slip effects
[slip length varying from several micrometers ($\mu$m) to several
hundreds $\mu$m] were observed on nanostructured superhydrophobic
surfaces\cite{Joseph2006,Choi2006,Lee2008,Karatay2013,Wu2014}.

\begin{figure}
  \includegraphics[bbllx=10pt,bblly=10pt,bburx=300pt,bbury=340pt,width=0.5\textwidth]{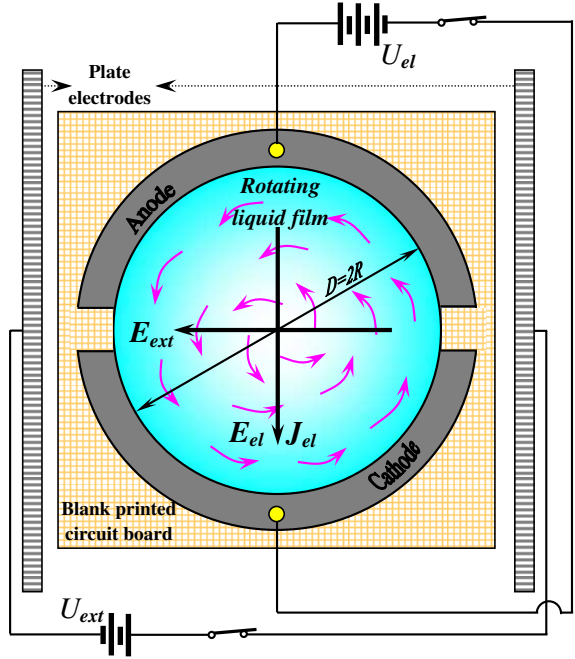}
\caption{(Color online) Schematic picture of the PLFM operated with
DC fields. The device consists of a two dimensional frame with two
graphite (or copper) electrodes (gray strips) on the sides for
electrolysis of the liquid film (blue-green zone). The radius and
diameter of the film are denoted, respectively, as $R$ and $D$. The
frame is made of an ordinary blank printed circuit board with a
circular (or square) hole at the center. The diameter of the hole
may vary from several centimeters to less than a millimeter.
Suspended liquid films as thin as hundreds of nanometers or less may
be created by brushing the liquid on the frame. The electric current
$\emph{\textbf{J}}_{el}$ (induced by electrolysis field
$\emph{\textbf{E}}_{el}$) and an external electric field
$\emph{\textbf{E}}_{ext}$ are produced by two circuits with voltage
$U_{el}$ and $U_{ext}$, respectively. $\emph{\textbf{E}}_{ext}$,
induced by two plates (striate strips) of a large capacitor, is
perpendicular to $\emph{\textbf{J}}_{el}$. When the magnitudes of
$\emph{\textbf{E}}_{el}$ and $\emph{\textbf{E}}_{ext}$ are above
threshold values, the film rotates, i.e., constitutes a motor. The
rotation direction obeys a right-hand rule, i.e.,
$\emph{\textbf{E}}_{ext}\times\emph{\textbf{J}}_{el}$. If the DC
electric sources (bold vertical lines in circuits) are replace by AC
ones, PLFM can also rotate in AC fields with the same frequencies.}
\label{Fig.1}       
\end{figure}

The goal of our study is to investigate dynamical properties of
PLFMs under slip boundary conditions. Such investigations are of
theoretical, experimental and technological significance. Firstly,
exploring the impact of interactions between liquids and solids on
EHD motions will advance our understanding of fluid mechanics.
Secondly, slippage on liquid-solid interfaces affects fluid
transportation in micro- and nano-fluidic systems:\cite{Bocquet2010}
large boundary slip can reduce hydrodynamic drag in micro- and nano-
channels\cite{Bocquet2010,Barratetal990102040614}, improving the
detection efficiency of the micro-fluidic chips, i.e., the study of
related mechanisms and laws is helpful to accelerate developments of
lab-on-a-chip technology. Fortunately, PLFMs open new ways to
explore aforementioned issues. We expect investigations on boundary
slippage to elucidate several experimental boundary phenomena of
various PLFMs types and to delineate optimizing methods for
realizing PLFM's technological applications in the lab-on-a-chip. In
this paper we focus on effects of boundary slip on rotation
characteristics of PLFMs.

The outline of this paper is as follows: In Sec. II, we present a
model for PLFMs with slip boundary conditions, and derive their
general solutions describing their EHD motions. In Sec. III, we
derive a series of specific characteristics of the DC and the AC
PLFMs, and compare these with experimental ones. Our conclusions we
present in Sec. IV. For convenience, the DC motor (DCM) and the AC
motor (ACM) denotations are used to represent the DC and the AC
PLFMs, respectively. We stress that in this paper we only
theoretically derive characteristics of the DCM and the ACM under
slip boundary conditions. We do not report any new experimental
data. All the experimental results cited in our paper were obtained
by different research groups and reported in the literature.

\section{Model of PLFMs with slip boundary conditions and its solutions}

Our models of the DCM and the ACM, developed in previous
publications\cite{Liuetal2011,Liuetal2012}, are based on the
assumption that a polar liquid film in an external electric field
can be depicted as a Bingham plastic fluid with an effective
electric dipole moment. Quantum field theoretical (quantum
electro-dynamic) aspects of polar
liquids\cite{Giudice1988,Sivasubramanian0506,Preparata199588,Sivasubramanian2001020309,Yinnon2009,Huang2009},
e.g., water, together with experimental results and their analyses,
e.g., of exclusion zone (EZ)
water\cite{Pollack2006to2013,Yinnon2014} and the Floating Water
Bridge\cite{Fuchs2010,Giudice2010}, underlie this assumption -- as
discussed in Sec. II of reference \cite{Liuetal2011} and Sec. II A
of reference \cite{Liuetal2012}. With our models we obtained
dynamical characteristics of the DCM and the ACM with no slip
boundary conditions, we successfully qualitatively and
quantitatively explained many experimental phenomena and we made a
series of predictions\cite{Liuetal2011,Liuetal2012}, e.g., we
predicted the EHD rotation threshold fields, the relation between
the rotation speed and the phase difference of the AC fields and the
vibration frequency of the ACM. Very recent experiments verified our
predictions pertaining to the EHD rotations and the plastic
vibrations of the ACM\cite{Amjadietal2013}.

Encouraged by our models' previous successes, we employ these to
study the slip-boundary effects on the EHD rotation properties of
the DC and the AC PLFMs. In our models, inspired by the
experimentally observed stable ring structure of the rotating
circular (or square) liquid film, we divided the liquid film into a
series of concentric cylindrical
discs\cite{Liuetal2011,Liuetal2012}, all of which obey the
rotational form of Newton's second
law. The dynamics equation, developed in references\cite{Liuetal2011} and \cite{Liuetal2012}, reads%
\begin{equation}
u_{t}=\frac{\mu }{\rho r^{2}}\left( r^{2}u_{rr}+ru_{r}-u\right) +\frac{%
\Delta\left(t \right)}{\rho r},\ \ \left( 0\leq r\leq R,\, t\geq
0\right) . \label{derlf}
\end{equation}%
Here $u_{t}$ denotes the first partial derivative of the linear
velocity $u\left(r,t\right)$ of the disc's rotation with respect to
time $t$; $u_{rr}$ and $u_{r}$ respectively represent the second and
the first partial derivative of $u\left(r,t\right)$ with respect to
the radius $r$, which is the average radius of a concentric
cylindrical disc from $r$ to $r+dr$; $\mu$, $\rho$ and $R=D/2$,
respectively, are the plastic viscosity, the density and the radius
of the liquid film; the driving
source of the EHD rotations of the liquid film at time $t$ is \cite{Liuetal2011}, %
\begin{equation}
\Delta\left(t \right)=\varepsilon _{0}\left( 1-1/\varepsilon
_{r}\right) E_{ext}\left(t \right)E_{el}\left(t \right)\sin
\theta_{EJ}-2\tau_{0}, \label{deltat}
\end{equation}%
where $\varepsilon _{0}$, $\varepsilon _{r}$ and $\tau_{0}$,
respectively, are the dielectric constant of the vacuum, the
relative dielectric constant and the yield stress of the liquid
film; $E_{ext}\left(t \right)$ and $E_{el}\left(t \right)$,
respectively, denote the magnitudes of the external electric field
$\emph{\textbf{E}}_{ext}$ and of the electrolysis electric field
$\emph{\textbf{E}}_{el}$ at time $t$; $\theta_{EJ}$ is the angle
between $\emph{\textbf{E}}_{ext}$ and $\emph{\textbf{E}}_{el}$.
Generally, $\theta_{EJ}=\pi/2$, i.e., $\emph{\textbf{E}}_{ext}$ is
perpendicular to $\emph{\textbf{E}}_{el}$, as shown in Fig. 1.

\begin{figure}
  \includegraphics[bbllx=5pt,bblly=10pt,bburx=590pt,bbury=130pt,width=1.0\textwidth]{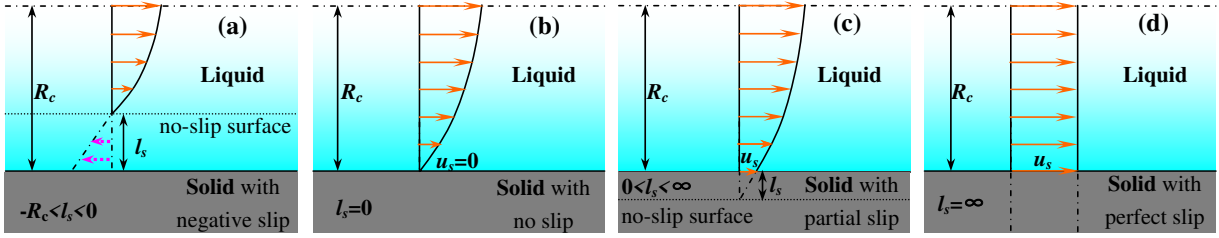}
\caption{(Color online) Schematic transverse cross-sections of an
infinite long cylindrical channel filled with liquid, with slip
boundary conditions described by different slip lengths $l_{s}$.
$R_{c}$ denotes the channel's radius. (a) For $-R_{c}<l_{s}<0$, the
liquid's linear velocity in the channel, i.e., $u_{c}\left(r,t
\right)$ (represented with orange arrows) as a function of $r$
quickly diminishes to zero in the liquid near the boundary if there
is negative slip at the liquid-solid interface. Pink dotted arrows
denote an imaginary reverse flow. (b) For $l_{s}=0$, $u_{c}\left(r,t
\right)$ gradually diminishes to zero near the boundary if there is
no slip at the solid-liquid interface, i.e., $u_{s}=0$. (c) For
$0<l_{s}<\infty$, when boundary slip occurs at the solid-liquid
interface, there is relative velocity between fluid flow and the
cylinder boundary, i.e., $u_{s}>0$. (d) For $l_{s}=\infty$, the
solid-liquid interface does not exert any resistance on the fluid,
i.e., $u_{c}\left(r,t \right)$ is independent of $r$ and
$u_{s}=u_{c}\left(r,t \right)$. The horizontal dash-dot lines and
the horizontal dotted lines represent the central line of channels
and the no-slip surfaces, respectively.}
\label{Fig.2}       
\end{figure}

In variation on our previous models, which assumed a no-slip
boundary condition [i.e., $u\left( r,t\right) |_{r=R}=0$], in the
slip-boundary model of this paper, Eq. (1) should satisfy two
boundary conditions and an initial condition: the disappearance of
the linear velocity at $r=0$, a nonzero slip velocity [i.e.,
$u_{s}=u\left( r,t\right) |_{r=R}$] of the linear velocity at
$r=R$, and the liquid film is static at $t=0$, i.e.,%
\begin{equation}
u\left( r,t\right) |_{r=0}=0,\ \ u_{s}=-l_{s}u_{r}|_{r=R}\ ,
\label{Bc}
\end{equation}%
and%
\begin{equation}
u\left( r,t\right) |_{t=0}=0.  \label{initialc}
\end{equation}
Here $u_{s}$ is the slip velocity of a fluid at the liquid-solid
interface, $u_{r}$ is the velocity gradient in a direction normal to
the surface and $l_{s}$ is the so-called slip length. Navier was the
first to define $l_{s}$;\cite{Navier1827} nowadays it customarily is
used to characterize the type of flow in
channels,\cite{Neto2005,Ranjith2013,Song2013} e.g., negative-slip,
no-slip, partial slip and perfect-slip flows in micro- or nano-
channels in lab-on-a-chip devices. For a transverse cross section of
an infinite long cylindrical channel, $l_{s}$ is defined as an
extrapolated distance relative to its wall where the tangential
velocity component vanishes (see Fig. 2c).\cite{Navier1827,Neto2005}
Negative-slip, no-slip, partial-slip and perfect-slip conditions are
described with different $l_{s}$ values (see Fig. 2): if
$-R_{c}<l_{s}<0$, with $R_{c}$ denoting the radius of the cylinder,
the flow is negative slip flow (i.e., locking
boundary)\cite{Song2013}, see Fig. 2a; if $l_{s}=0$, the flow is
stick flow (i.e., no slip boundary), see Fig. 2b; if $l_{s}=\infty$,
the flow is plug flow (i.e., perfect slip boundary), see Fig. 2d;
intermediate values of $l_{s}$ represent partial slip flow, see Fig.
2c. We stress that the boundary zone with the negative linear
velocity in Fig. 2a does not represent the existence of a reverse
flow; it may be considered as a static liquid zone;\cite{Song2013}
its thickness in low permeability porous media is often denoted
``the boundary layer thickness".\cite{Song2013}

\begin{figure}
  \includegraphics[bbllx=20pt,bblly=30pt,bburx=330pt,bbury=340pt,width=0.5\textwidth]{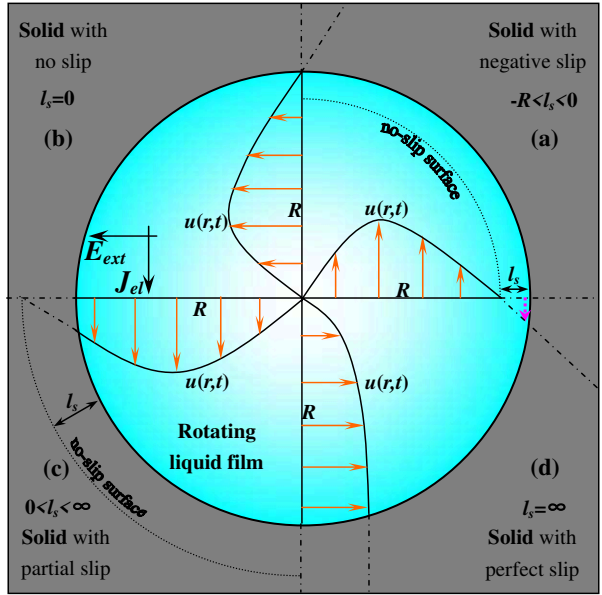}
\caption{(Color online) Schematic linear velocity's profiles of the
slip boundary conditions, with different slip lengths $l_{s}$ in a
rotating liquid film. (a) $-R<l_{s}<0$ , negative-slip boundary; (b)
$l_{s}=0$, no-slip boundary; (c) $0<l_{s}<\infty$ , partial-slip
boundary; (d) $l_{s}=\infty$, perfect-slip boundary.}
\label{Fig.3}       
\end{figure}

With the PLFM's suspended film corresponding to a $\sim10^2$nm thick
slice of a cylindrical channel, in our model we adopt the
aforementioned definitions of $l_{s}$. The film's schematic profiles
with above-defined boundary conditions are plotted in Fig. 3. Its
denotations are the same as those used in Fig. 2.

Eqs. (\ref{derlf}), (\ref{Bc}) and (\ref{initialc}) describe the EHD
rotations of PLFMs under the above mentioned four different
slip-boundary conditions described with different $l_{s}$ values.
The general solutions of our model's mixed problem can be obtained
by the method of eigenfunctions\cite{Mathews1971}. Assuming $u\left(
r,t\right) =R_{f}\left( r\right)T\left( t\right)$, inserting it into
the homogeneous equation of Eq. (\ref{derlf}) and Eq. (\ref{Bc}),
respectively, we obtain the eigenvalues problem%
\begin{subequations}  \label{EVP}
\begin{numcases}{}
R^{\prime \prime }_{f}\left( r\right) +R^{\prime }_{f}\left(
r\right) /r+\left(
\lambda _{n}^{2}-1/r^{2}\right) R_{f}\left( r\right) =0,  \label{EVPA}\\
R_{f}\left( 0\right) =0,\ \  R_{f}\left( R\right) +l_{s}R^{\prime
}_{f}\left( R\right) =0,    \label{EVPB}
\end{numcases}
\end{subequations}
where separation of variables method\cite{Mathews1971} was used to
introduce eigenvalues $\lambda _{n}$ . The eigenfunctions of Eq.
(\ref{EVP}), depicting the spatial modes in the general solutions of
Eq. (\ref{derlf}), are a series of the ordinary Bessel functions of
order one: $J_{1}\left( \lambda_{n}r\right)$, $r\subseteq[0,R]$,
$n=1,2,\cdots$. Obviously, $J_{1}\left( \lambda_{n}r\right)$ satisfy
the first boundary condition in Eq. (\ref{EVPB}) and the
corresponding eigenvalues $\lambda _{n}=\kappa _{n}/R$ are
determined by the transcendental
equation%
\begin{equation}
J_{1}\left( \kappa _{n}\right) +\frac{l_{s}}{D}\kappa _{n}\left[
J_{0}\left( \kappa _{n}\right) -J_{2}\left( \kappa _{n}\right)
\right] =0,  \label{transcendental-equation}
\end{equation}
which is a natural result when $J_{1}\left( \lambda_{n}r\right)$
satisfy the second boundary condition in Eq. (\ref{EVPB}). Here
$J_{0}\left( \xi\right)-J_{2}\left( \xi\right)=2J^{\prime}_{1}\left(
\xi\right)$ was used, $J_{0}\left( \xi\right)$ and $J_{2}\left(
\xi\right)$, respectively, represent the Bessel functions of orders
0 and 2. For a given $k=l_{s}/D$ , $\kappa _{n}$ and $\lambda _{n}$
can be obtained numerically from Eq.
(\ref{transcendental-equation}). The values of $\lambda _{n}$,
varying with different $k$, determine the behaviors of $J_{1}\left(
\lambda_{n}r\right)$, which reflect the spatial modes of the
rotating liquid film. Therefore the rotation properties of PLFMs
with the slip boundary conditions depend on $k$, but do not depend
on the size of liquid film $D$ and the slip length $l_{s}$.

From Eqs. (\ref{EVP}) and (\ref{transcendental-equation}), one may
prove that the aforementioned Bessel functions $J_{1}\left(
\lambda_{n}r\right)$, in variation on the functions obtained in our
previous works\cite{Liuetal2011,Liuetal2012}, should obey the
following orthogonality relations (see Appendix A)
\begin{equation}
\int\nolimits_{0}^{R}rJ_{1}\left( \lambda _{m}r\right) J_{1}\left(
\lambda _{n}r\right) dr=\frac{R^{2}}{2}\left[ 1+\frac{1}{\lambda
_{n}^{2}}\left( \frac{1}{l_{s}^{2}}-\frac{1}{R^{2}}\right) \right]
J_{1}^{2}\left( \lambda _{n}R\right) \delta _{mn}, \label{OrthR}
\end{equation}
where $\delta _{mn}=1$ when $m=n$, $\delta _{mn}=0$ when $m\neq n$.
Since the above Bessel function series is a complete orthogonal
system, the general solution to Eq. (\ref{derlf}) and the last term
$f\left( r,t\right) =\Delta \left( t\right)/\rho r$ in Eq.
(\ref{derlf}) may
be expanded by them in generalized Fourier series, i.e.,%
\begin{equation}
u\left( r,t\right) =\sum\limits_{n=1}^{\infty }T_{n}\left( t\right)
J_{1}\left( \lambda _{n}r\right), \label{general-solution}
\end{equation}
and%
\begin{equation}
f\left( r,t\right) =\Delta \left( t\right) \sum\limits_{n=1}^{\infty
}C_{n}J_{1}\left( \lambda _{n}r\right), \label{DSGFS}
\end{equation}
where
\begin{equation}
C_{n}=\frac{2\kappa _{n}\left[ 1-J_{0}\left( \kappa _{n}\right) \right] }{%
R\rho \left[ \kappa _{n}^{2}+\left( R^{2}/l_{s}^{2}-1\right) \right]
J_{1}^{2}\left( \kappa _{n}\right) }. \label{Cn}
\end{equation}

Inserting Eq. (\ref{general-solution}) into Eq. (\ref{Bc}), one
finds that the first boundary condition of Eq. (\ref{derlf}) is
satisfied automatically and the second one yields Eq.
(\ref{transcendental-equation}). Inserting Eqs.
(\ref{general-solution}) and (\ref{DSGFS}) into  Eqs. (\ref{derlf})
and
(\ref{initialc}), respectively, we have%
\begin{equation}
T_{n}^{\prime }\left( t\right) +a_{n}T_{n}\left( t\right)
=C_{n}\Delta \left( t\right) \label{Tn(t)DE}
\end{equation}
and $T_{n}\left( 0\right) =0$, here $a_{n}=\frac{\mu }{\rho }%
\frac{\kappa _{n}^{2}}{R^{2}}$. The general solution to Eq.
(\ref{Tn(t)DE}) is%
\begin{equation}
T_{n}\left( t\right) =e^{-a_{n}t}[ \int C_{n}\Delta \left( t\right)
e^{a_{n}t}dt+Q_{n}] \label{Tn(t)},
\end{equation}
where the constant $Q_{n}$ is determined by $T_{n}\left( 0\right)
=0$.

Eqs. (\ref{general-solution}) and (\ref{Tn(t)}) present our model's
general solutions for PLFMs under slip boundary conditions. The
linear velocity distribution of rotating PLFMs is given by Eq.
(\ref{general-solution}), in which the spatial modes and the time
factors are respectively expressed by $J_{1}\left(
\lambda_{n}r\right)$ satisfying Eqs.
(\ref{transcendental-equation}), (\ref{OrthR}) and by Eq.
(\ref{Tn(t)}). The corresponding angular velocity reads%
\begin{equation}
\omega \left( r,t\right) =u\left( r,t\right) /r. \label{AV}
\end{equation}

\section{Results and discussion}

PLFMs can work perfectly with many different crossing electric
fields, e.g., DC\cite{Amjadietal2009,Shirsavaretal2011,Liuetal2011},
AC\cite{Amjadietal2009,Liuetal2012}, square-wave\cite{Liuetal2013}
and other type\cite{Liuetal2012-2}. In this study, we present the
boundary slip effects on the rotation properties of PLFMs operated
with DC and AC fields, and compare these with experimentally
observed results.

\subsection{DCM with slip boundary conditions}

According to Eq. (\ref{deltat}), for DCM $\Delta \left(
t\right)$ is a constant, i.e., independent of time $t$:%
\begin{equation}
\Delta ^{dc}=\Delta \left( t\right)=\varepsilon _{0}\left(
1-1/\varepsilon _{r}\right) E_{ext}E_{el}\sin \theta _{EJ}-2\tau
_{0}. \label{DeltaDC}
\end{equation}
On inserting Eq. (\ref{DeltaDC}) into Eq. (\ref{Tn(t)}), we obtain
the time factors describing the rotation evolution of the DCM:
\begin{equation}
T_{n}^{dc}\left( t\right) =\frac{C_{n}\Delta ^{dc}}{a_{n}}\left(
1-e^{-a_{n}t}\right).  \label{Tn(t)DC}
\end{equation}
From Eqs. (\ref{general-solution}), (\ref{Cn}) and (\ref{Tn(t)DC}),
we obtain the linear velocity of the DCM%
\begin{equation}
u^{dc}\left( r,t\right) =\sum\limits_{n=1}^{\infty
}C_{n}^{dc}J_{1}\left( \frac{\kappa _{n}r}{R}\right) \left(
1-e^{-a_{n}t}\right),  \label{DCLV}
\end{equation}
where
\begin{equation}
C_{n}^{dc}=\frac{2R\Delta ^{dc}}{\mu }\frac{1-J_{0}\left( \kappa
_{n}\right) }{\kappa _{n}\left[ \kappa _{n}^{2}+\left(
R^{2}/l_{s}^{2}-1\right) \right] J_{1}^{2}\left( \kappa _{n}\right)}
\end{equation}
The corresponding angular velocity is given by Eqs. (\ref{AV}) and
(\ref{DCLV}).

\begin{figure}
  \includegraphics[bbllx=10pt,bblly=10pt,bburx=480pt,bbury=360pt,width=0.9\textwidth]{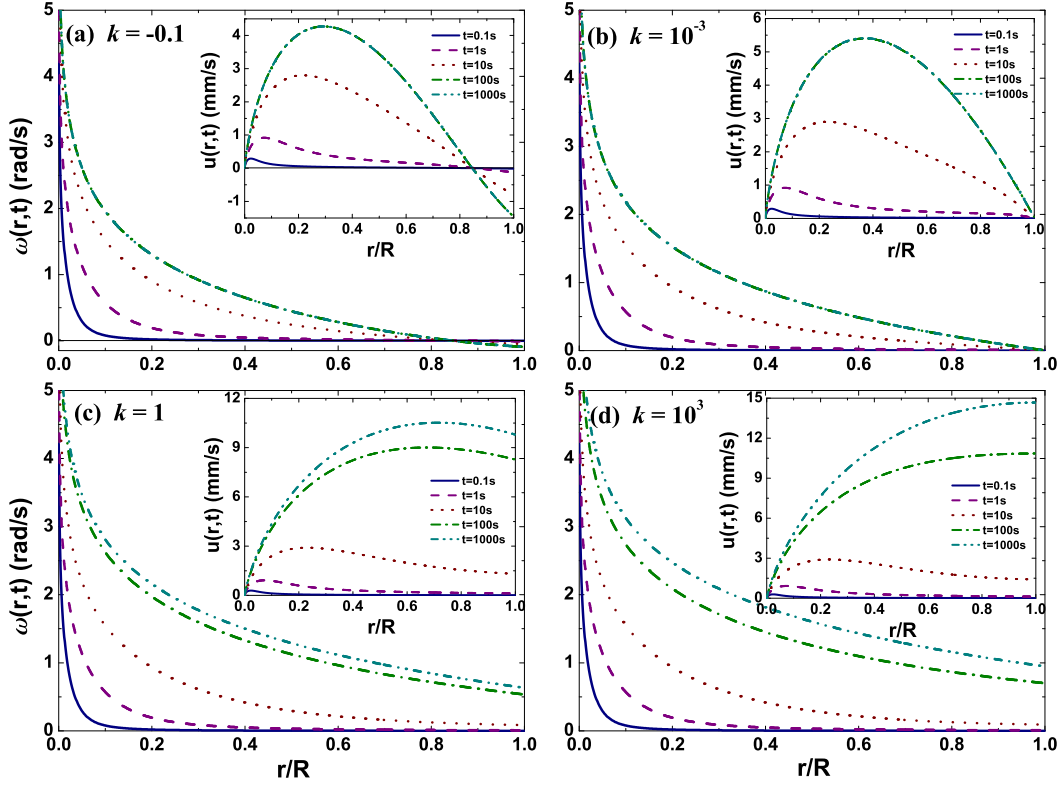}
\caption{(Color online) The profiles of the angular velocity of the
DCM with four different boundary conditions represented by different
values of $k$ at different times: (a) $k=-0.1$; (b) $k=10^{-3}$; (c)
$k=1$; (d) $k=10^{3}$. The insets show the profiles of the
corresponding linear velocity of each figure.}
\label{Fig.4}       
\end{figure}

To illuminate the dynamical characteristics of the DCM for different
$k$ values, we adopt the experimental parameters of the exemplary
extensively measured and theoretically investigated
DCM\cite{Amjadietal2009,Shirsavaretal2011,Liuetal2011,Liuetal2012},
i.e.: $\varepsilon _{0}=8.85\times 10^{-12}$ F$\cdot$m$^{-1}$,
$\varepsilon _{r}=80$, $E_{ext}U_{el}\sin \theta_{EJ} =7.2\times
10^{6}$ V$^{2}\cdot$m$^{-1}$, $D=2R=3.1\times 10^{-2}$ m, $\rho
=10^{3}$ kg$\cdot$m$^{-3}$, $\mu =10^{-3}$ Pa$\cdot$s and its
derived $\tau _{0}=6.77\times 10^{-5}$ Pa. With these parameters, we
investigated characteristics of the angular and linear velocities
dependencies on $k$. We found that by setting $k=-0.1$, $10^{-3}$,
$1$ or $10^{3}$, the DCM rotates under, respectively, negative-slip,
approximately-no-slip, partial-slip and approximately-perfect-slip
boundary conditions, as discernible from Fig. 4: This figure depicts
the profiles of  our computed angular and linear velocities at
different times -- by drawing the intersection points of the tangent
lines to the curves with $t=1000$s at the point $R$ and the
horizontal axis in the insets of Fig. 4, it is noticeable that Figs.
4a-d are consistent with the four slip boundary cases presented by
Figs. 3a-d, respectively. To illustrate this more clearly, we plot
the profiles of the steady rotation angular velocities of the DCM in
Fig. 5a. The main rotation characteristics of the DCM, observable
from Figs. 4 and 5a, are:

(i) The points near the center of the film, independent of the
boundary conditions, start to rotate earlier than those farther away
from it. This derived result is in full agreement with the
experimental one.\cite{Amjadietal2009,Shirsavaretal2011}

\begin{figure}
  \includegraphics[bbllx=10pt,bblly=10pt,bburx=470pt,bbury=360pt,width=0.9\textwidth]{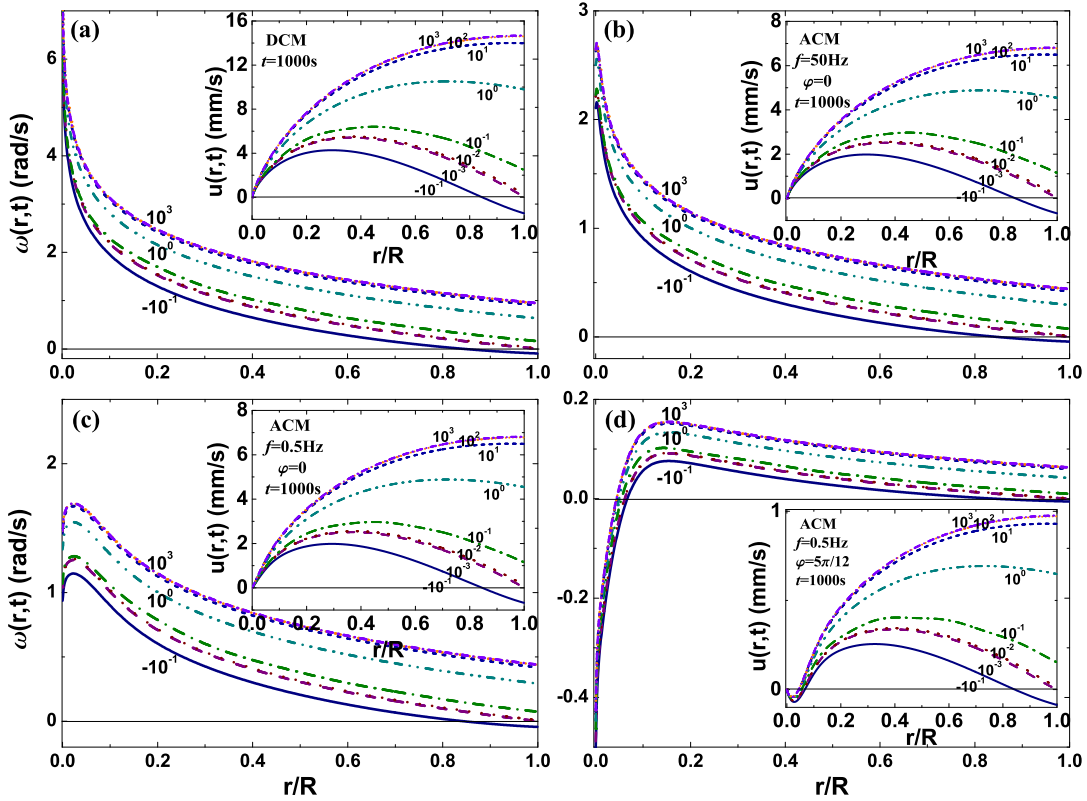}
\caption{(Color online) The profiles of the angular velocity of the
steady rotating DCM and ACM with different $k$ values. $k$ varies
from $-0.1$ to $10^{3}$. (a) DCM; (b) ACM with $f=50$Hz and
$\varphi=0$; (c) ACM with $f=0.5$Hz and $\varphi=0$; (d) ACM with
$f=0.5$Hz and $\varphi=5\pi/12$. The insets show the profiles of the
corresponding linear velocity of each figure. Obviously, the
rotation properties of the DCM depend on $k$, while those of the ACM
are associated with $k$ as well as these depend on the frequencies
and on the phase difference of the AC fields.}
\label{Fig.5}       
\end{figure}

(ii) For any $k$, the angular velocities decrease with increasing
$r$, i.e., these are smaller at points further away from the film's
center -- see Fig. 4. This prediction too fully agrees with the
experimental ones.\cite{Amjadietal2009,Shirsavaretal2011}

(iii) As $k$ increases, the angular velocity of the steady rotation
grows gradually and its decay rate with $r$ deceases slowly.
Experiments capable of verifying this prediction have not yet been
reported and are called for.

(iv) The linear velocity's distribution mainly depends on the ratio
$k$, but not on the particular values of $R$ and $l_{s}$, i.e., it
exhibits significant different characteristics as $k$ increases. For
some $k$ values, the linear velocity distributions are qualitatively
consistent with experimental ones, as we discuss in the next
paragraphs a-d.

a. For $-1/2<k<0$, the DCM rotates under a negative slip boundary
condition, i.e., locking boundary -- see Fig. 4a which shows that as
$r$ increases from zero (i.e., from the center of the film), the
rotation's linear velocity increases quickly from zero to a maximum
and then decreases slowly to zero, even to a negative value as $r$
increases further.

As mentioned in Sec. II, the zone around a border with negative
linear velocity (see pink dashed arrows in Figs. 2a and 3a) may be
considered as a zone with static liquid. Simulation with the lattice
Boltzmann method show that a strong solid-fluid attraction may
result in a small negative slip length\cite{Zhang2004}. Molecular
dynamics simulations show that the no-slip or locking boundary
conditions correspond to ordered liquid structures close to the
solid walls leading to zero and negative slip lengths,
respectively\cite{Xu2007}. Strong solid-fluid attractions and
ordered liquid structures close to the solid walls have been
observed in numerous recent experiments: A wide zone of
long-ordered, high-viscosity and liquid-crystal-phase water adjacent
to objects like hydrophilic membranes, reactive metal sheets,
biological tissues, optical fibers, gels with charged or uncharged
surfaces has been observed\cite{Pollack2006to2013}.

Combining the experimental and simulations' results cited in the
last paragraph with our theoretical ones, we predict that the DCM
may exhibit a physics picture given by Fig. 4a when motor's frames
are made of hydrophilic materials. This conjecture is partly
supported by the experimental results of rotating liquid crystal
films driven by crossing DC electric fields:\cite{Shirsavaretal2012}
Figure 4a in this paper is qualitatively consistent with Figs. 2d
and 5 in reference \cite{Shirsavaretal2012}. Since water and liquid
crystals have different micro-structures and properties, further
experiments are called for to verify the above prediction and more
detailed theoretical investigations are needed.

b. For $k$ a small positive number, e.g. $0<k\leq 10^{-2}$, the DCM
rotates under an approximately-no-slip condition -- Figs. 4b and 5a
show that the rotation properties of the DCM are almost the same as
those obtained on assuming no slip boundary condition, see Figs. 7
and 8 of Ref. \cite{Liuetal2011}. The linear velocity has its
maximum at $R/e$, with $e$ representing the irrational and
transcendental constant approximately equal to 2.71828, and the
angular velocity is a decreasing function of the radius $r$. This
result shows that the slip boundary effect may be ignored when the
slip length $l_{s}$ is much less than the size of the liquid film.
It also indirectly proves that our model is reasonable.

c. For $10^{-2}<k<\sim10$, the DCM rotates under partial slip
condition -- we can discern this result by drawing the intersection
points of the tangent lines to the curves at the point $r/R=1$ and
the horizontal axis in the insets of Figs. 4 and 5a, and
subsequently compare with Fig. 3c. As $k$ increases above $10^{-2}$,
the maximum value of the linear velocity of the steady rotating
liquid film increases and its location moves from $R/e$ to $R$, as
shown in Fig. 5a.

For the magnitudes of the external electric fields and solid-liquid
border parameters, adopted above for our exemplary DCM, it is
reasonable to assume $l_{s}$ is constant. Thus, Fig. 5a indicates
that the linear and angular velocities will increase as the size of
the liquid film decreases. It is also noteworthy that for $k=1$ and
$t=1000$s, the maximum value of the linear velocity locates around
$2R/e$, the angular velocity decays slowly with increasing $r$ and
it has a large value at the boundary (see Fig. 4c). These properties
are qualitatively consistent with the experimental ones exhibited by
the rotating Benzonitrile film (see Fig. 3c in reference
\cite{Shirsavaretal2011}). Hence investigating the slip boundary
effects facilitates understanding some experimental results
unexplainable with our previous model\cite{Liuetal2011}, which did
not treat slip boundaries.

d. For sufficient large $k$, e.g., $k\geq10$ (see Fig. 5a),  the DCM
rotates under approximately-perfect-slip boundary condition -- we
can obtain this result by the same method mentioned in case c. The
angular velocity is a decreasing function of the radius $r$ (see
Fig. 4d) and the linear velocity of steady rotation is an increasing
function of the radius $r$ (see the inset in Fig. 4d). To the best
of our knowledge, hitherto, there are no corresponding experimental
results for verifying these theoretical ones.

\subsection{ACM with slip boundary conditions}

\begin{figure}
  \includegraphics[bbllx=10pt,bblly=10pt,bburx=470pt,bbury=360pt,width=0.9\textwidth]{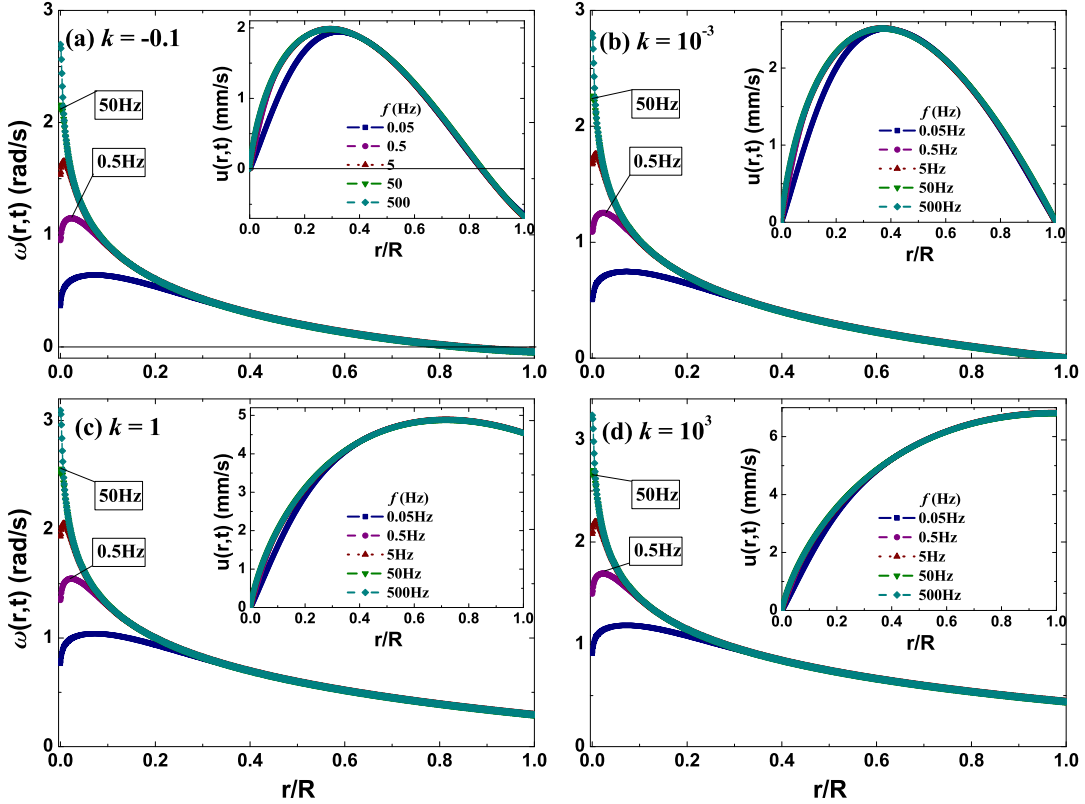}
\caption{(Color online) The profiles of the angular velocity of the
steady rotating ACM with different values of $f$ for $\varphi=0$ at
time $t=1000$s. $f$ varies from 0.05 Hz to 500Hz. (a) $k=-0.1$; (b)
$k=10^{-3}$; (c) $k=1$; (d) $k=10^{3}$. The insets show the profiles
of the corresponding linear velocity of each figure. Obviously, the
rotation properties around the film's center depend on $f$, while
those near the film's boundary are associated with $k$. As $f$
decreases, the angular velocities vary from a monotonically
decreasing function to a first increasing then decreasing function.
As $k$ increases, the ACM subsequently exhibits negative-slip,
no-slip, partial-slip, perfect-slip behaviors.}
\label{Fig.6}       
\end{figure}

Experimental\cite{Amjadietal2009,Amjadietal2013} and our previously
published theoretical results\cite{Liuetal2012} show: when the
crossing AC fields have the same frequencies, the ACM exhibits
rotation characteristics similar to those of the DCM; AC fields with
different frequencies can merely induce vibrations not rotation. Our
detailed previously published predictions on ACM's rotation and
vibration characteristics\cite{Liuetal2012} are now fully confirmed
by experiments\cite{Amjadietal2013}, with the exception of some
details of the elastic vibration -- a model based on our previous
published one, with the improvement of assuming the film to be an
elastic Bingham fluid, explains these details\cite{Amjadietal2013}.
In the model presented in this paper, we did not include the
aforementioned improvement, i.e., we do not expect it to describe
all vibration properties of ACMs. Still, based on its previous
success in correctly describing ACMs' rotation properties, we
conjecture our model can elucidate the effect of slip boundaries on
the ACM rotations.

\begin{figure}
  \includegraphics[bbllx=10pt,bblly=10pt,bburx=470pt,bbury=360pt,width=0.9\textwidth]{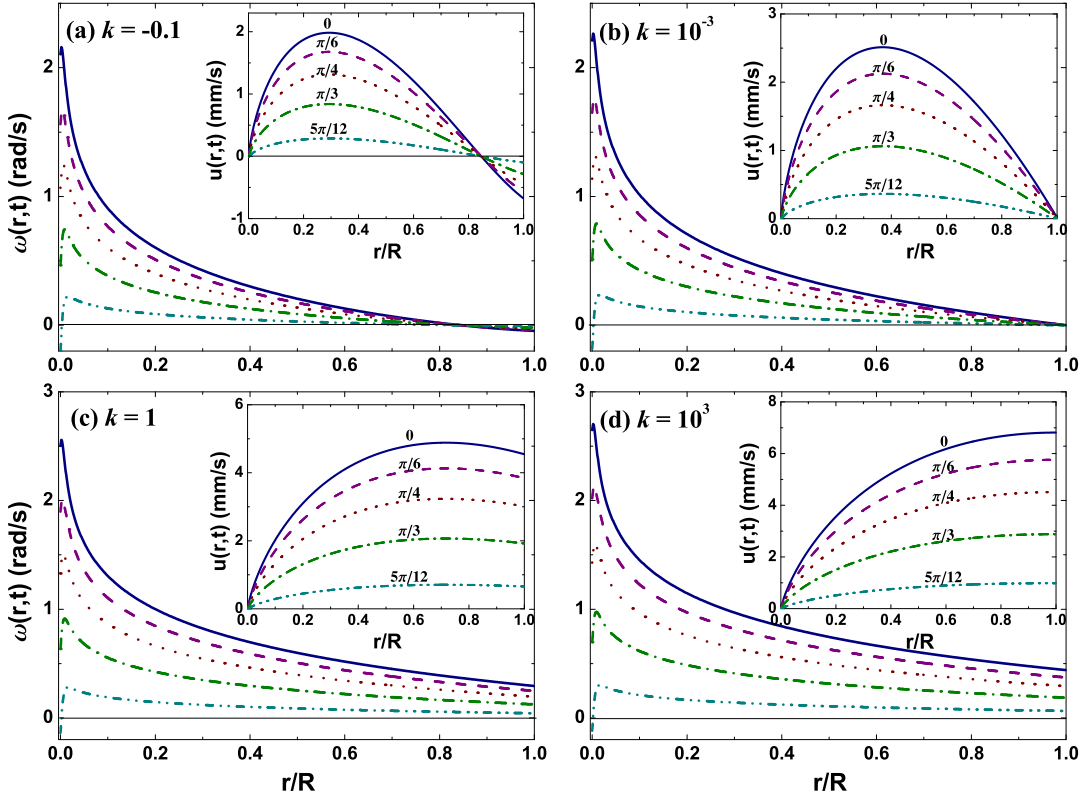}
\caption{(Color online) The profiles of the angular velocity of the
rotating ACM with different $k$ values for various values of
$\varphi$ ($\varphi=0$, $\pi/6$, $\pi/4$, $\pi/3$, $5\pi/12$) at
time $t=1000$s: (a) $k=-0.1$; (b) $k=10^{-3}$; (c) $k=1$; (d)
$k=10^{3}$. The frequencies of the AC fields are 50Hz. The insets
show the profiles of the corresponding linear velocity of each
figure. Obviously, the angular velocities are decreasing functions
of the radius. On comparing curves in this figure with those with
$t=1000$s in Fig. 4, for the corresponding $k$ values, one finds
that the ACM and the DCM exhibit similar characteristics when AC
fields' frequencies are large (e.g., $f=50$Hz). As $\varphi$
increases, the rotation speed decreases gradually.}
\label{Fig.7}       
\end{figure}

\begin{figure}
\includegraphics[bbllx=10pt,bblly=10pt,bburx=470pt,bbury=360pt,width=0.9\textwidth]{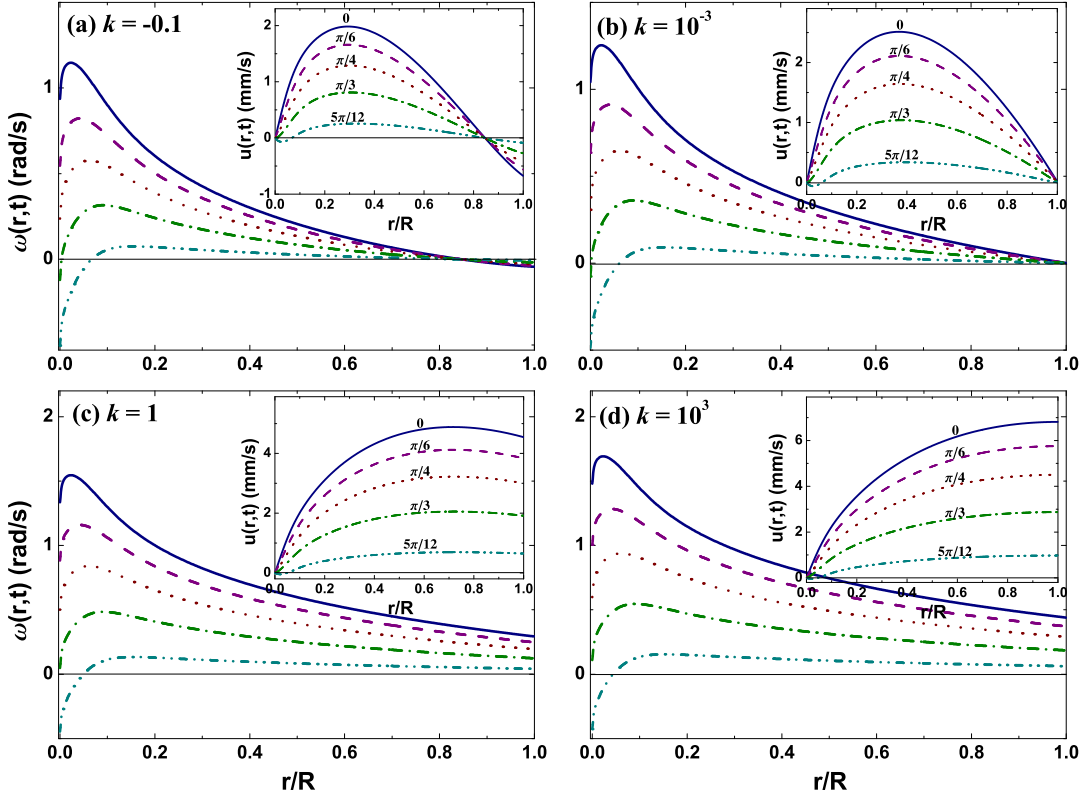}
\caption{(Color online) The profiles of the angular velocity of the
rotating ACM with different $k$ values for various values of
$\varphi$ ($\varphi=0$, $\pi/6$, $\pi/4$, $\pi/3$, $5\pi/12$) at
time $t=1000$s: (a) $k=-0.1$; (b) $k=10^{-3}$; (c) $k=1$; (d)
$k=10^{3}$. The insets show the profiles of the corresponding linear
velocity of each figure. The frequencies of the  AC fields are
0.5Hz. As $\varphi$ increases, the rotation speed not only decreases
gradually, but also its maximum moves away from the center of the
film.}
\label{Fig.8}       
\end{figure}

For the analysis we define the alternating external electric field
and electrolysis voltage, respectively, as $E_{ext}\left(
t\right)=E_{0}\sin \omega^{ac} t$ and $U_{el}\left(
t\right)=E_{el}\left( t\right)D=U_{0}\sin \left( \omega^{ac}
t+\varphi \right)$. Here $E_{0}$ and $U_{0}$, respectively, denote
the amplitudes of the electric field and the voltage,
$\omega^{ac}=2\pi f$ represents their angular frequencies, $\varphi
$ is the initial phase of the electrolysis voltage and it also
represents the phase difference between the AC fields. From Eq.
(\ref{deltat}),
we obtain the source driving the rotation of the ACM (as explicitly shown in Ref. \cite{Liuetal2012} -- Sec. III B)%
\begin{equation}
\Delta ^{ac}\left( t \right)=B_{c}+B_{t}=\left( b^{ac}\cos \varphi
-2\tau _{0}\right) -b^{ac}\cos \left( 2\omega^{ac} t+\varphi
\right), \label{DeltaAC}
\end{equation}
where $b^{ac}=\frac{\varepsilon _{0}\left( 1-1/\varepsilon
_{r}\right) E_{0}U_{0}\sin \theta _{EJ}}{4R}$. Eq. (\ref{DeltaAC})
is suitable for studying the rotation of the ACM only if $B_{c}>0$,
which means the active torque should first destroy the plastic
structure of the liquid film before it can induce the film to
rotate. Inserting Eq. (\ref{DeltaAC}) into Eq. (\ref{Tn(t)}), we
obtain the time factors
describing the evolution of the ACM, i.e.,%
\begin{equation}
T_{n}^{ac}\left( t\right) =C_{n}^{acBc}\left( 1-e^{-a_{n}t}\right)
+C_{n}^{acBt}\cos \gamma _{n}\left[ \cos \left( \varphi -\gamma
_{n}\right) e^{-a_{n}t}-\cos \left( 2\omega^{ac} t+\varphi -\gamma
_{n}\right) \right] , \label{TntAC}
\end{equation}
where $C_{n}^{acBc}=C_{n}B_{c}/a_{n}$,
$C_{n}^{acBt}=C_{n}b^{ac}/a_{n}$, $\gamma _{n}=\arctan \left(
2\omega^{ac} /a_{n}\right)$. By employing Eqs.
(\ref{general-solution}) and (\ref{TntAC}), we have
\begin{eqnarray}
u^{ac}\left( r,t\right)  &=&\sum\limits_{n=1}^{\infty
}C_{n}^{acBc}J_{1}\left( \frac{\kappa _{n}r}{R}\right) \left(
1-e^{-a_{n}t}\right) \nonumber \\
&&+\sum\limits_{n=1}^{\infty }C_{n}^{acBt}J_{1}\left( \frac{%
\kappa _{n}r}{R}\right) \cos \gamma _{n}\left[ \cos \left( \varphi
-\gamma
_{n}\right) e^{-a_{n}t}-\cos \left( 2\omega^{ac} t+\varphi -\gamma _{n}\right) %
\right]. \label{LVAC}
\end{eqnarray}%
The corresponding angular velocity is given by Eqs. (\ref{AV}) and
(\ref{LVAC}). Eq. (\ref{LVAC}) indicates that for the ACM, rotation
of the film comprises not only rotation modes but also plastic
vibration modes. Our calculations, reported in the following
paragraphs, show the contributions of these two different types of
modes vary with the magnitudes, the frequencies and the phase
difference of the AC fields. Thus, the dynamical characteristics of
the ACM with slip boundary conditions depend on $k$ as well as on
the aforementioned variables. Obviously, Eqs. (\ref{DeltaAC}) and
(\ref{LVAC}) show that the rotation speed increases as the AC
fields' magnitudes increase, because higher fields induce larger
average active torque in the film. The aforementioned results agree
well with the experimental ones: measurements show that in the case
of phase differences, vibration and rotation exist simultaneously
and by increasing the magnitude of the electric fields, the ratio
between vibration and rotation velocities changes. In high electric
fields the rotation dominates, while in lower fields the vibration
dominates\cite{Amjadietal2013}.

To illuminate in detail the dynamical characteristics of the ACM for
different $k$, $f$ and $\varphi$ values, we adopt in the expression
for $b^{ac}$ [employed in Eq.(\ref{DeltaAC})] $E_{0}U_{0}\sin
\theta_{EJ} =7.2\times 10^{6}$ V$^{2}\cdot$m$^{-1}$, i.e., the same
value as that used in our previous analyses \cite{Liuetal2012}; for
the other parameters we adopt the values of the DCM of Section III
A. With Eqs. (\ref{AV}) and (\ref{LVAC}), we computed numerous
profiles of the angular velocity of the steady rotating ACM for
different $k$, $f$ and $\varphi$ values and in depth analyzed these.
Exemplary cases, which depict the main trends, we plot in Figs.
5b-d, 6-8.

From Figs. 5b-d -- which exhibit the profiles of the angular and
linear velocity of the steady rotating ACM with different $k$ values
($k$ varies from $-0.1$ to $10^3$) for the three cases of $f=50$Hz
and $\varphi=0$, of $f=0.5$Hz and $\varphi=0$ and of $f=0.5$Hz and
$\varphi=5\pi/12$ -- we learn:

(i) The boundary dynamical behaviors of the ACM, just as those of
the DCM studied earlier in Sec. III A, are determined by $k$: On
comparing Figs. 5b-d with Fig. 3, it is discernible that as $k$
increases ACM subsequently exhibits ``negative-", ``no-",
``partial-" and ``perfect-" slip behaviors. To the best of our
knowledge, neither experimental nor computational data have been
reported that can verify Fig.5's predictions, i.e., creation of such
data is called for.

(ii) For given AC fields' magnitudes, the contributions of rotation
modes and plastic vibration modes depends on $f$ and $\varphi$:

a. For high $f$ and small $\varphi$, e.g., $f\geq50$Hz and
$\varphi=0$, the ACM exhibits characteristics similar to those of
the DCM: All the angular velocities are almost monotonically
decreasing functions of $r$ and all the linear velocities increase
with increasing $k$ -- as can be seen by comparing Fig. 5a with 5b.
These theoretical results agree well with the experimental ones: the
measured dynamical characteristics of the rotation of the ACM, for
frequencies in the range of 50Hz up to 40kHz, are the same as those
of the DCM\cite{Amjadietal2009}.

b. For low $f$ and small $\varphi$, e.g., $f\leq0.5$Hz and
$\varphi=0$, the ACM and the DCM exhibit different properties. The
angular velocities of the ACM do not monotonically decrease with
$r$, as do those of the DCM --  as can be seen by comparing Fig. 5a
with 5c. The ACM's angular velocities increase first and then
decrease as $r$ enhances. Moreover, our computations show that the
location of the maximum of the angular velocity moves away from the
center of the liquid film as the AC fields' frequencies decrease,
see Fig. 6. The underlying physical reason for this dependency of
the ACM's angular velocities on $r$ is that the contributions of the
plastic vibration modes to the linear velocity increase as AC
fields' frequencies decrease. When the frequencies are higher, the
minor and faster plastic vibrations, arising from the second term of
Eq. (\ref{DeltaAC}), can not induce a macroscopic flow in the liquid
film in each half period of $T_{1}=1/(2f)$. As $f$ decreases, the
plastic vibration modes' ability to produce a macroscopic reverse
flow in the liquid film strengthens and leads to the velocities
presented in Figs. 5c, 6 and 8.  These theoretical results agree
well with the numerical ones given in Fig. 6 of reference
\cite{Amjadietal2013}, which show that as frequencies increase, the
plastic vibration gets weaker.

c. For large $\varphi$ and low $f$ values, e.g., $\varphi=5\pi/12$
and $f=0.5$Hz (see Fig. 5d), the ACM exhibit an interesting
phenomenon: the region near the center of the film and that near the
border may rotate in opposite directions. The underlying reason is
that as $\varphi$ increases, $B_{c}=\left( b^{ac}\cos \varphi -2\tau
_{0}\right)$ in Eq. (\ref{DeltaAC}) decreases. This leads to a
diminishment of the rotation modes' contributions to the EHD motions
in the ACM, while those of vibration modes become more distinct. To
the best of our knowledge, hitherto, there are no corresponding
experimental results for verifying these theoretical ones.

Detailed effects of $\varphi$ on the rotation characteristics of the
ACM we present in Figs. 7 and 8. These figures show that the
rotation speed of the ACM decreases as $\varphi$ increases.
Moreover, these show when $f$ is high enough, the positional
deviation of the maximum of the angular velocity from the center of
the film is negligible with increasing $\varphi$, see Fig. 7.
However, when $f$ is small, this deviation becomes significant as
$\varphi$ increases: When $\varphi$ is large enough, the region near
the center of the film and that near the border may rotate in
opposite directions (see curves with $\varphi=5\pi/12$ in Fig. 8).
The aforementioned theoretical results agree with the experimental
and calculated ones\cite{Amjadietal2013}: Experimental results show
that average tangential velocities of an ACM, with a phase
difference of $\varphi=5\pi/12=75^{0}$, frequency of 41Hz and
$E_{ext0}E_{el0} =2.2\times 10^{8}$ V$^{2}\cdot$m$^{-2}$, may be
positive or negative -- see Fig. 3 in reference
\cite{Amjadietal2013}; Calculated results of shear rates show that
the average values of the plastic shear rate and the rotatory shear
rate are opposite in sign -- see the bottom picture of Fig. 5 in
reference \cite{Amjadietal2013}.

\section{Summary and conclusions}

By introducing slip boundary conditions in the DCM and ACM models
(developed by us in our previous
publications),\cite{Liuetal2011,Liuetal2012} we investigated their
rotation characteristics for different slip boundary types. To the
best of our knowledge, our study is the first to theoretically
derive slip boundary effects on these PLFM's EHD.

Our computations with our slip boundary DCM and ACM models, for
small boundary slip length $l_{s}$, i.e., approximately-no-slip
condition, respectively, recover the dynamical characteristics of
the DCM and the ACM with no-slip boundary condition. Our
computations also present the rotation properties of these motors
with negative-slip, partial-slip and perfect-slip conditions. These
properties shed light on several experimentally observed rotation
characteristics of the DCM and ACM at their borders. Our
theoretically derived properties agree with the existing
experimental ones and those obtained with numerical techniques.
Additional experiments, required for verifying our remaining derived
properties which are specified in Sec. III A and B, are called for.

One of our central theoretical results is that the rotation speed
distributions are independent of the absolute values of the $l_{s}$
and the liquid film's diameter $D$, but are associated with their
ratio $k=l_{s}/D$:

(i) As $k$ ($k>-1/2$) increases, DCM and ACM subsequently exhibit
rotation characteristics under ``negative-", ``no-", ``partial-" and
``perfect-" slip boundary conditions;

(ii) For the DCM, $k$ affects the magnitude of its linear velocity:
for small $k$ values, this magnitude first increases and then
decreases with the radius $r$ of the liquid film at its steady
rotation; as $k$ increases, this magnitude increases and the
location of its maximum approaches the film's boundary. As to the
effects of $k$ on the DCM's angular velocity: for any $k$, it always
decays with $r$; as $k$ increases, its decay rate slows down, and
this results in a nonzero angular velocity at the border of the
liquid film.

(iii) For the ACM, its EHD motions depends on $k$, as well as on the
magnitude, frequencies and phase difference of the AC fields;
moreover these motions comprise rotation modes and plastic vibration
modes. At high frequencies, the ACM exhibits similar rotation
characteristics to the DCM with the same $k$. At low-frequency, its
rotation properties distinctly differ from those of the DCM, because
the plastic vibration may weaken the rotation around the film's
center -- as a result the region near the center of the film and
that near the border even may rotate in opposite directions when the
phase difference is large enough.

As to desirable future research, we note that the slip length
$l_{s}$ was introduced into our model to study the effects of slip
boundary on the dynamical properties of the DCM and the ACM.
However, we did not study any factors affecting the values of
$l_{s}$, such as the solid-fluid potential interactions, surface
roughness, wettability\cite{Neto2005}, the presence of gaseous
layers\cite{Vinogradova2011,Karatay2013} and dipole moment of polar
liquids\cite{Cho2004}. As such, future research on factors affecting
$l_{s}$ is called for. In particular, because as mentioned in Sec.
I, for example, the nano-structured super-hydrophobic surface can
produce large slip length in the solid-liquid
interface\cite{Joseph2006,Choi2006,Lee2008,Karatay2013,Wu2014}. We
expect large $l_{s}$ to accelerate mixing effects of PLFMs -- for
PLFMs operating at high rotational speeds induced by identical
external electric fields, large $l_{s}$ seems to be of special
importance for optimizing the micro-mixer investigated in one of our
previous studies\cite{Liuetal2013}.

We conclude with mentioning: the study of water slippage on
different interfaces has always been an important and even a hot
issue\cite{Neto2005,Lauga2007}; our study indicates that fabricating
PLFMs with different hydrophilic or hydrophobic frames offers
opportunities for studying the solid-liquid interfacial slippage; we
expect such a study to  deepen our understanding of the dynamical
properties of the PLFMs; we anticipate that such investigation
ultimately will lead to delineation of those film characteristics
required for optimizing the PLFMs.

\appendix\section{Orthogonality relations among $J_{1}\left( \lambda_{n}r\right)$}

The first differential equation in Eq. (\ref{EVP}) may be simplified
as
\begin{equation}
\frac{d}{dr}\left( r\frac{dR_{f}}{dr}\right) +\left( \lambda ^{2}r-\frac{1}{%
r}\right) R_{f}=0. \label{A1}
\end{equation}
If $m\neq n$, $\lambda _{m}\neq \lambda _{n}$, inserting
$J_{1}\left( \lambda_{n}r\right)$ and $J_{1}\left(
\lambda_{m}r\right)$ into Eq. (\ref{A1}), respectively, we obtain
\begin{equation}
\frac{d}{dr}\left( r\frac{dJ_{1}\left( \lambda _{n}r\right)
}{dr}\right) +\left( \lambda _{n}^{2}r-\frac{1}{r}\right)
J_{1}\left( \lambda _{n}r\right) =0, \label{A2}
\end{equation}
\begin{equation}
\frac{d}{dr}\left( r\frac{dJ_{1}\left( \lambda _{m}r\right)
}{dr}\right) +\left( \lambda _{m}^{2}r-\frac{1}{r}\right)
J_{1}\left( \lambda _{m}r\right) =0. \label{A3}
\end{equation}
Let $J_{1}\left( \lambda_{m}r\right)$ multiply Eq. (\ref{A2}),
$J_{1}\left( \lambda_{n}r\right)$ multiply Eq. (\ref{A3}), and the
former minus the latter, then calculate the integral of them from
$0$ to $R$, we obtain
\begin{equation}
\left( \lambda _{n}^{2}-\lambda _{m}^{2}\right)
\int\nolimits_{0}^{R}rJ_{1}\left( \lambda _{m}r\right)
J_{1}\left(\lambda _{n}r\right) dr=J_{1}\left( \lambda _{n}R\right)
\lambda _{m}RJ_{1}^{^{\prime }}\left( \lambda _{m}R\right)
-J_{1}\left( \lambda _{m}R\right) \lambda _{n}RJ_{1}^{^{\prime
}}\left( \lambda _{n}R\right). \label{A4}
\end{equation}
Since $\lambda _{n}$ and $\lambda _{m}$ should satisfy Eq.
(\ref{transcendental-equation}), we obtain
\begin{equation}
\lambda _{n}RJ_{1}^{^{\prime }}\left( \lambda _{n}R\right) =-\frac{R}{l_{s}}%
J_{1}\left( \lambda _{n}R\right),\ \lambda _{m}RJ_{1}^{^{\prime }}\left( \lambda _{m}R\right) =-\frac{R}{l_{s}}%
J_{1}\left( \lambda _{m}R\right). \label{A5}
\end{equation}
On inserting Eq. (\ref{A5}) into Eq. (\ref{A4}), one finds for
$m\neq n$,
\begin{equation}
\int\nolimits_{0}^{R}rJ_{1}\left( \lambda _{m}r\right) J_{1}\left(
\lambda _{n}r\right) dr=0. \label{A6}
\end{equation}
Next, let us discuss the case $m=n$. When $m\rightarrow n$, i.e.,
when $\lambda _{m}\rightarrow\lambda _{n}$, if $\lambda _{n}$
satisfies the first equation in Eq. (\ref{A5}), $\lambda _{m}$ does
not satisfy the second equation in Eq. (\ref{A5}), i.e., $\lambda _{m}RJ_{1}^{^{\prime }}\left( \lambda _{m}R\right) \neq -\frac{R}{l_{s}}%
J_{1}\left( \lambda _{m}R\right)$. As $m\rightarrow n$, from Eq.
(\ref{A4}) we have
\begin{equation}
\int\nolimits_{0}^{R}rJ_{1}\left( \lambda _{m}r\right) J_{1}\left(
\lambda
_{n}r\right) dr=\underset{\lambda _{m}\rightarrow \lambda _{n}}{\lim }\frac{%
J_{1}\left( \lambda _{n}R\right) \lambda _{m}RJ_{1}^{^{\prime
}}\left( \lambda _{m}R\right) +\frac{R}{l_{s}}J_{1}\left( \lambda
_{m}R\right) J_{1}\left( \lambda _{n}R\right) }{\lambda
_{n}^{2}-\lambda _{m}^{2}} . \label{A7}
\end{equation}
Applying L Hospital rule to the right hand of Eq. (\ref{A7}), i.e.,
deducing the derivative of the numerator and denominator with
respect to $\lambda _{m}$, respectively, we obtain%
\begin{equation}
\int_{0}^{R}rJ_{1}^{2}\left( \lambda _{n}r\right) dr=-\frac{%
R^{2}J_{1}\left( \lambda _{n}R\right) }{2\lambda _{n}^{2}}\left[
\lambda _{n}J_{1}^{^{\prime }}\left( \lambda _{n}R\right) /R+\lambda
_{n}^{2}J_{1}^{^{^{\prime \prime }}}\left( \lambda _{n}R\right)
+\lambda _{n}J_{1}^{^{\prime }}\left( \lambda _{n}R\right)
/l_{s}\right] . \label{A8}
\end{equation}
From Eq. (\ref{A2}) and the first equation of (\ref{A5}), we obtain
$\lambda _{n}^{2}J_{1}^{^{^{\prime \prime }}}\left( \lambda
_{n}R\right) =-\lambda _{n}J_{1}^{^{\prime }}\left( \lambda
_{n}R\right) /R-\left( \lambda _{n}^{2}-1/R^{2}\right) J_{1}\left(
\lambda _{n}R\right) $, and $\lambda _{n}J_{1}^{^{\prime }}\left(
\lambda _{n}R\right) =-J_{1}\left( \lambda _{n}R\right) /l_{s}$.
Inserting them into Eq. (\ref{A8}), we obtain
\begin{equation}
\int_{0}^{R}rJ_{1}^{2}\left( \lambda _{n}r\right)
dr=\frac{R^{2}}{2}\left[
1+\frac{1}{\lambda _{n}^{2}}\left( \frac{1}{l_{s}^{2}}-\frac{1}{R^{2}}%
\right) \right] J_{1}^{2}\left( \lambda _{n}R\right) . \label{A9}
\end{equation}
Combining Eq. (\ref{A6}) with Eq. (\ref{A9}), we obtain the
orthogonality relations given by Eq. (\ref{OrthR}).

\begin{acknowledgments}
Our work has been supported by National Natural Science Foundation
of China (No. 11302118), Natural Science foundation of Shandong
Province (No. ZR2013AQ015) and the Science Foundation of Qufu Normal
University (No. BSQD2012053). Tamar Yinnon expresses her
appreciation for Prof. A. M. Yinnon's continuous support and
encouragement.
\end{acknowledgments}

\newpage
\begin{center}
\textbf{Figure Captions}
\end{center}

Fig. 1: (Color online) Schematic picture of the PLFM operated with
DC fields. The device consists of a two dimensional frame with two
graphite (or copper) electrodes (gray strips) on the sides for
electrolysis of the liquid film (blue-green zone). The radius and
diameter of the film are denoted, respectively, as $R$ and $D$. The
frame is made of an ordinary blank printed circuit board with a
circular (or square) hole at the center. The diameter of the hole
may vary from several centimeters to less than a millimeter.
Suspended liquid films as thin as hundreds of nanometers or less may
be created by brushing the liquid on the frame. The electric current
$\emph{\textbf{J}}_{el}$ (induced by electrolysis field
$\emph{\textbf{E}}_{el}$) and an external electric field
$\emph{\textbf{E}}_{ext}$ are produced by two circuits with voltage
$U_{el}$ and $U_{ext}$, respectively. $\emph{\textbf{E}}_{ext}$,
induced by two plates (striate strips) of a large capacitor, is
perpendicular to $\emph{\textbf{J}}_{el}$. When the magnitudes of
$\emph{\textbf{E}}_{el}$ and $\emph{\textbf{E}}_{ext}$ are above
threshold values, the film rotates, i.e., constitutes a motor. The
rotation direction obeys a right-hand rule, i.e.,
$\emph{\textbf{E}}_{ext}\times\emph{\textbf{J}}_{el}$. If the DC
electric sources (bold vertical lines in circuits) are replace by AC
ones, PLFM can also rotate in AC fields with the same frequencies.
\bigskip

Fig. 2: (Color online) Schematic transverse cross-sections of an
infinite long cylindrical channel filled with liquid, with slip
boundary conditions described by different slip lengths $l_{s}$.
$R_{c}$ denotes the channel's radius. (a) For $-R_{c}<l_{s}<0$, the
liquid's linear velocity in the channel, i.e., $u_{c}\left(r,t
\right)$ (represented with orange arrows) as a function of $r$
quickly diminishes to zero in the liquid near the boundary if there
is negative slip at the liquid-solid interface. Pink dotted arrows
denote an imaginary reverse flow. (b) For $l_{s}=0$, $u_{c}\left(r,t
\right)$ gradually diminishes to zero near the boundary if there is
no slip at the solid-liquid interface, i.e., $u_{s}=0$. (c) For
$0<l_{s}<\infty$, when boundary slip occurs at the solid-liquid
interface, there is relative velocity between fluid flow and the
cylinder boundary, i.e., $u_{s}>0$. (d) For $l_{s}=\infty$, the
solid-liquid interface does not exert any resistance on the fluid,
i.e., $u_{c}\left(r,t \right)$ is independent of $r$ and
$u_{s}=u_{c}\left(r,t \right)$. The horizontal dash-dot lines and
the horizontal dotted lines represent the central line of channels
and the no-slip surfaces, respectively.
\bigskip

Fig. 3: (Color online) Schematic linear velocity's profiles of the
slip boundary conditions, with different slip lengths $l_{s}$ in a
rotating liquid film. (a) $-R<l_{s}<0$ , negative-slip boundary; (b)
$l_{s}=0$, no-slip boundary; (c) $0<l_{s}<\infty$ , partial-slip
boundary; (d) $l_{s}=\infty$, perfect-slip boundary.
\bigskip

Fig. 4: (Color online) The profiles of the angular velocity of the
DCM with four different boundary conditions represented by different
values of $k$ at different times: (a) $k=-0.1$; (b) $k=10^{-3}$; (c)
$k=1$; (d) $k=10^{3}$. The insets show the profiles of the
corresponding linear velocity of each figure.
\bigskip

Fig. 5: (Color online) The profiles of the angular velocity of the
steady rotating DCM and ACM with different $k$ values. $k$ varies
from $-0.1$ to $10^{3}$. (a) DCM; (b) ACM with $f=50$Hz and
$\varphi=0$; (c) ACM with $f=0.5$Hz and $\varphi=0$; (d) ACM with
$f=0.5$Hz and $\varphi=5\pi/12$. The insets show the profiles of the
corresponding linear velocity of each figure. Obviously, the
rotation properties of the DCM depend on $k$, while those of the ACM
are associated with $k$ as well as these depend on the frequencies
and on the phase difference of the AC fields.
\bigskip

Fig. 6: (Color online) The profiles of the angular velocity of the
steady rotating ACM with different values of $f$ for $\varphi=0$ at
time $t=1000$s. $f$ varies from 0.05 Hz to 500Hz. (a) $k=-0.1$; (b)
$k=10^{-3}$; (c) $k=1$; (d) $k=10^{3}$. The insets show the profiles
of the corresponding linear velocity of each figure. Obviously, the
rotation properties around the film's center depend on $f$, while
those near the film's boundary are associated with $k$. As $f$
decreases, the angular velocities vary from a monotonically
decreasing function to a first increasing then decreasing function.
As $k$ increases, the ACM subsequently exhibits negative-slip,
no-slip, partial-slip, perfect-slip behaviors.
\bigskip

Fig. 7: (Color online) The profiles of the angular velocity of the
rotating ACM with different $k$ values for various values of
$\varphi$ ($\varphi=0$, $\pi/6$, $\pi/4$, $\pi/3$, $5\pi/12$) at
time $t=1000$s: (a) $k=-0.1$; (b) $k=10^{-3}$; (c) $k=1$; (d)
$k=10^{3}$. The frequencies of the AC fields are 50Hz. The insets
show the profiles of the corresponding linear velocity of each
figure. Obviously, the angular velocities are decreasing functions
of the radius. On comparing curves in this figure with those with
$t=1000$s in Fig. 4, for the corresponding $k$ values, one finds
that the ACM and the DCM exhibit similar characteristics when AC
fields' frequencies are large (e.g., $f=50$Hz). As $\varphi$
increases, the rotation speed decreases gradually.
\bigskip

Fig. 8: (Color online) The profiles of the angular velocity of the
rotating ACM with different $k$ values for various values of
$\varphi$ ($\varphi=0$, $\pi/6$, $\pi/4$, $\pi/3$, $5\pi/12$) at
time $t=1000$s: (a) $k=-0.1$; (b) $k=10^{-3}$; (c) $k=1$; (d)
$k=10^{3}$. The insets show the profiles of the corresponding linear
velocity of each figure. The frequencies of the  AC fields are
0.5Hz. As $\varphi$ increases, the rotation speed not only decreases
gradually, but also its maximum moves away from the center of the
film.

\end{document}